# Star Forming Galaxy Models: Blending Star Formation into TREESPH

J. Christopher Mihos and Lars Hernquist[1]
*Board of Studies in Astronomy and Astrophysics,*
*University of California, Santa Cruz, CA 95064*
*hos@lick.ucsc.edu, lars@lick.ucsc.edu*

## ABSTRACT

We have incorporated star formation algorithms into a hybrid $N$-body /smoothed particle hydrodynamics code (TREESPH) in order to describe the star forming properties of disk galaxies over timescales of a few billion years. The models employ a Schmidt law of index $n \sim 1.5$ to calculate star formation rates, and explicitly include the energy and metallicity feedback into the ISM. Modeling the newly formed stellar population is achieved through the use of hybrid SPH/young star particles which gradually convert from gaseous to collisionless particles, avoiding the computational difficulties involved in creating new particles. The models are shown to reproduce well the star forming properties of disk galaxies, such as the morphology, rate of star formation, and evolution of the global star formation rate and disk gas content. As an example of the technique, we model an encounter between a disk galaxy and a small companion which gives rise to a ring galaxy reminiscent of the Cartwheel (AM 0035–35). The primary galaxy in this encounter experiences two phases of star forming activity: an initial period during the expansion of the ring, and a delayed phase as shocked material in the ring falls back into the central regions.

*Subject headings:* galaxies:evolution, galaxies:individual (AM 0035–035), galaxies:starburst, methods:numerical, stars:formation

---

## 1. Introduction

The study of galaxy formation and evolution involves a complicated range of physical processes which includes gravitational dynamics, hydrodynamics, the effects of star formation, and the formation and evolution of stellar populations. These processes are nonlinear and closely interlinked. As an illustration, consider the formation of a bar in a disk galaxy which gives rise to a nuclear burst of star formation. As the bar grows in strength, the gas in the bar forms a ridge of high density material on the leading edge of the bar (Roberts, Huntley, & van Albada 1979; Norman 1988). The subsequent torque on the gas from the slightly trailing stellar bar leads to a rapid inflow of disk gas (Combes 1985; Hernquist & Mihos 1994). The increased gas densities in the central regions of the disk can fuel a strong central burst of star formation, depleting the central gas and creating a new population of young stars (Mihos, Richstone, & Bothun 1992; Mihos & Hernquist 1994a). The energy input from the stellar winds and supernovae associated with the starburst may trigger galactic superwinds and slow or halt the inflow of gas (Heckman, Armus, & Miley 1990). The subsequent evolution of the starburst population can leave signatures in the photometric and dynamical properties of the galaxy for some time. The close connection between these differing physical processes make full evolutionary models of disk galaxies difficult to construct.

Recent advances in computer technology and numerical methods have made possible detailed modeling of the gravitational dynamics and hydrodynamics of disk galaxies. The development of $N$-body codes which employ a hierarchical tree structure (e.g., Barnes & Hut 1986; Hernquist 1987) allows for a favorable $O(N \log N)$ scaling of computing time without placing restrictions on the spatial resolution or global geometry. As a result, $N$-body realizations of disk galaxies with $N \sim 10^5 - 10^6$ are now feasible, which reduces the compromising effects of two body scattering and $\sqrt{N}$ noise in the potential. The dynamics of disk galaxies may now be followed for many rotation periods with spatial resolutions of a few hundred parsecs. To model the ISM of disk galaxies, smoothed particle hydrodynamics (SPH; e.g., Lucy 1977; Gingold & Monaghan 1977), provides an effective way of representing the disk gas as individual fluid elements which carry the hydrodynamic properties of the gas. The fully Lagrangian approach of SPH allows it to be integrated in a straightforward manner into $N$-body algorithms, resulting in a detailed gravitational and hydrodynamical description of the evolution of disk galaxies. A successful integration of these two schemes is TREESPH (Hernquist & Katz 1989), a hybrid $N$-body/SPH code which has been used to model the evolution of many complex systems, including disk galaxy mergers (e.g., Barnes & Hernquist 1991, 1992; Hernquist 1989; Hernquist & Weil 1992; Hernquist & Mihos 1994) and galaxy formation processes (Katz & Gunn 1991; Katz 1992).

However, further investigation into the evolution of disk galaxies requires that star formation be incorporated into the modeling techniques. Such an integration has proved difficult for a number of reasons. First, the detailed physics which describe star formation are still not well understood, on either small (parsec) or large (kiloparsec) size scales. The numerical method by which ISM material is converted into stars, therefore, must be based either on simplistic gravitational collapse models or on empirical observations of nearby disk galaxies. Second, the computational task of converting gas to stellar material requires either that new particles be created to represent the young stars (e.g., Katz 1992, Navarro & White 1993), or a complete conversion of a gas particle into a star particle (Summers 1993). The former method is extremely costly in terms of computational resources, while the latter sets limitations on the mass resolution of a star forming event.

Motivated by a desire to model star formation and the triggering of starburst activity in interacting and merging galaxies, we have developed a technique that describes star formation processes in disk galaxies without placing restrictions on the mass resolution (other than that set by $N$) or resulting in inefficient allocation of computational resources. In this paper, we describe this method for incorporating star formation into TREESPH in a manner that is both finely resolved (in mass and time) and computationally feasible. Our approach results in a good description of star formation in disk galaxies over timescales of a few billion years, typical of the timescales involved in galaxy collisions and mergers. The actual star forming "law" is based on observations of nearby star forming galaxies, and is well-described by a Schmidt law with index $n \sim 1.5$ (Schmidt 1959). This choice of star formation law is easily changed in order to investigate the influence of a number of parameters on star formation, such as the role of threshold densities, gas velocity dispersion, or thermal energy.

The layout of this paper is as follows. In §2, we give a brief description of TREESPH, followed by a more detailed description of the star forming algorithms. We test these algorithms in §3 on two galaxy models: an isolated disk galaxy evolving in isolation, and a collision reminiscent of the Cartwheel ring galaxy. We summarize the modeling technique and outline areas of interest in §4.

## 2. Numerical Technique

Many approaches exist for modeling the gravitational and hydrodynamical evolution of disk galaxies. We wish to use these techniques to model the evolution of interacting and merging galaxies; for these systems, treecodes offer the best solution to calculating the gravitational forces, as they place no constraints on the symmetry or spatial resolution of the system being modeled. Similarly, to model the hydrodynamical evolution of the ISM in these galaxies, Lagrangian smoothed particle hydrodynamics (SPH) is well-suited to these problems – by representing the gas as discrete fluid elements which carry the hydrodynamical properties, no restrictions on spatial resolution or symmetry are imposed. Furthermore, the Lagrangian nature of SPH allows it to be easily integrated into $N$-body algorithms, providing for an efficient and elegant numerical approach to combined gravitational and hydrodynamical evolution. One such hybrid scheme is TREESPH (Hernquist & Katz 1989), which combines SPH into an $N$-body treecode, making it ideal for studying galaxy formation and evolution processes. Using TREESPH as a starting point, we incorporate into the numerical technique star formation prescriptions which describe star formation, gas depletion, and energy and metallicity injection into the ISM. We begin the description of the approach with a brief summary of the original TREESPH code, and then discuss star formation algorithms in detail.

### 2.1. TREESPH

To calculate the gravitational force acting on the $N$ particles representing the system, a hierarchical tree structure is employed, in which the particles are organized into progressively larger structures, or "nodes." Associated with each node are the mass, center of mass, and quadrupole moments of the particles contained within it. The force on a particle from a distant group of particles is then calculated using the information from the node containing the distant particles. If higher accuracy is required, the descendent nodes which more finely resolve the distant group of particles are examined. Using this technique, the computational time spent on each particle scales as $O(\log N)$, rather than the $O(N)$ scaling of traditional $N$-body methods. Gravitational forces are softened using a cubic spline of softening length $\epsilon$, which converges to the Keplerian formula exactly outside $r > 2\epsilon$ (Hernquist & Katz 1989). Particles may each be assigned their own softening lengths, allowing for better resolution of dense substructure. Pairwise forces are calculated using the mean of the individual softening lengths. With typical choices for accuracy tolerance and timestep, energy and angular momentum are conserved to much better than a percent.

In TREESPH, the gaseous component of the system is represented by discrete fluid elements which sample the smoothed hydrodynamic and thermodynamic properties of the gas. The particles are evolved using Euler's equation:

$$\frac{d\mathbf{r}_i}{dt} = \mathbf{v}_i, \qquad (1a)$$

$$\frac{d\mathbf{v}_i}{dt} = -\frac{1}{\rho_i}\nabla P_i + \mathbf{a}_i^{\rm visc} - \nabla \Phi_i, \qquad (1b)$$

where $\Phi_i$ is the gravitational potential, $P_i$ is the pressure, and $\mathbf{a}_i^{\rm visc}$ is an artificial viscosity term used to capture shocks in the flow. A conventional form of artificial viscosity is employed, with parameters $\alpha = 0.5$ and $\beta = 1.0$ (Hernquist & Katz 1989). The SPH particles have their individual timesteps chosen to satisfy a modified version of the Courant condition (Monaghan 1992).

To complete the hydrodynamic modeling of the ISM, TREESPH allows for different equations of state for the gas. In general, an ideal gas law may be used in conjunction with heating and cooling terms in the energy equation to describe the full thermal evolution of the gas (e.g., Hernquist & Katz 1989; Hernquist 1989; Barnes & Hernquist 1991; Katz 1992). In practice, however, an isothermal equation of state is usually used for modeling the ISM in disk galaxies. Due to limitations imposed by finite mass resolution, TREESPH is unable to accurately describe a true multi-phase ISM (e.g., McKee & Ostriker 1977); in order to suppress the formation of dense clumps of gas, therefore, radiative cooling is inhibited below a cutoff temperature of $\sim 10^4$ K. Owing to the short radiative cooling timescale of the ISM, most of the gas therefore resides very close to this cut-off temperature (Barnes & Hernquist 1991, 1994), resulting in little difference between models employing an isothermal equation of state and those employing a more "realistic" treatment of the gas microphysics. By choosing an isothermal equation of state, the interpretation of the results is simplified and the complexity of the calculation is reduced, with little sacrifice in reliability.

## 2.2. Star Formation Algorithms

Incorporating star formation into numerical simulations involves two major conceptual steps: a determination of the star formation rate, and a description of the effects of this star formation on the surrounding physical system. These two steps form a complex regulated cycle in the dynamics of the ISM which need to be handled concurrently in any numerical model. The star formation rate in a region of space is dependent on the physical conditions in the local ISM, while these physical conditions in turn depend on the energy input from the young stars embedded in the ISM. Our approach to incorporating this feedback cycle between the ISM and star forming regions is to model the star formation rate based on empirical relations appropriate for nearby disk galaxies, then constrain the feedback in such a way as to achieve a regulated, steady-state system. While it is not clear that descriptions of star formation derived from quiescent disk galaxies can be universally applied in the regime of starbursts and galaxy mergers, by using these descriptions as a starting point our models can determine whether the extreme star forming properties of these systems can be described by extending the disk star formation process into such violent regimes.

Ideally, the local star formation rate in a volume of space should be based on a detailed physical description of the star formation process. However, the relevant physics is poorly understood on the mass and size scales typical of our calculations. On small scales, smaller than a few parsecs, gravitational collapse models such as Jeans instability (Binney & Tremaine 1987) may provide a plausible description of star formation, although complications surely arise from effects associated with magnetic fields, the metallicity of the ISM, and rotational angular momentum of the collapsing cloud (Shu, Adams, & Lizano 1987). However, on the mass scales currently resolvable in simulations, typically $\sim 10^6$ M$\odot$, comparable to giant molecular clouds, Jeans collapse is surely an inappropriate description of star formation – while subunits in GMCs may be Jeans unstable, the clouds as a whole are probably stable (Scoville & Sanders 1987). On these larger mass scales, descriptions based on gravitational instability such as the Toomre instability criterion (Toomre 1963; Kennicutt 1989) may be more appropriate, in which the dynamics are governed by the local mass density and velocity dispersion. Such criteria, however, are based on gravitational instabilities in an isolated, rotating disk, and are not useful models for star formation in a rapidly changing gravitational field, such as in galaxy interactions and mergers, or during galaxy formation. Given the lack of a well-founded physical description for star formation at the resolution of our models, we turn to empirical star formation rates based on observations of star forming disks. With this approach, we are able to reproduce the star forming properties of disk galaxies without relying on physical laws such as the Jeans and Toomre criteria applied in inappropriate regimes.

Observations of nearby disk galaxies suggest that, on the size scale of hundreds of parsecs to a kiloparsec, the star formation rate per unit volume may be parameterized as a function of the local gas density, by a form of the "Schmidt law" (Schmidt 1959):

$$\rho_{\rm SFR} = C \times \rho_{\rm gas}^n. \qquad (2)$$

The best choice for $n$ is not well-determined, but several considerations argue for values in the range n $\sim$ 1–2. Observational determinations of $n$ based on star formation tracers such as H$\alpha$ emission or young star counts show a wide scatter but generally fall in the range $1 \lesssim n \lesssim 2.5$ (e.g., Berkhuijsen 1977; Guibert, Lequeux, & Viallefond 1978; Kennicutt 1989 and references therein). Furthermore, studies of spiral structure in nearby galaxies show that small increases in gas density in spiral arms are accompanied by relatively larger increases in star formation activity (Lord & Young 1990), implying $n > 1$. However, the fact that disk galaxies show a nearly constant rate of star formation over several billion years (Kennicutt 1983; Gallagher, Hunter, & Tutukov 1984) suggests that $n$ cannot be very large; otherwise disks would show a rapidly declining star formation rate with age. Finally, simple physical considerations argue for moderate values of $n$, at least as a first approximation. Star formation triggered by cloud collisions in the ISM (e.g., Larson 1969; Scoville, Sanders, &

Clemens 1986) yields $n \sim 1.5 - 2$, similar to values derived for parametrizations of simple gravitational instability mechanisms (Madore 1977; Larson 1987, 1988). Therefore, while clearly a vast simplification, a Schmidt law of moderate $n$ seems to be a reasonable parametrization of star formation on the kiloparsec size scale (although see Vasquez & Scalo 1989 for an alternate view).

With these considerations in mind, we choose to parametrize star formation in our models using a modified Schmidt law. Retaining the Lagrangian character of SPH, we describe the star formation rate in an SPH particle by

$$\frac{\dot{M}_{\text{gas}}}{M_{\text{gas}}} = C_{sfr} \times \rho_{\text{gas}}^{\frac{1}{2}}, \quad (3)$$

where $\dot{M}_{\text{gas}}$ is the star formation rate, $M_{\text{gas}}$ is the particle mass, and $\rho_{\text{gas}}$ is the smoothed estimate of the local gas density. This form of the star formation law yields a good representation of a Schmidt law of index $n \sim 1.5$ when averaged over volume (see §3.1). The normalization constant $C_{sfr}$ in equation 3 is chosen such that an isolated disk galaxy forms stars at roughly a rate of 1 $M_\odot$ yr$^{-1}$, similar to the global SFR in nearby disk galaxies (Kennicutt 1983).

Several authors have proposed modifications of the Schmidt law to include effects due to radial cutoffs in star formation (Lacey & Fall 1983, 1985), gas density thresholds (Kennicutt 1989) or galactic shear (Wyse & Silk 1989). We note that all of these modifications are motivated by gravitational instability arguments which are valid only for linear perturbations in thin rotating disks. The strong, rapidly changing tidal fields and three dimensional geometry which describe galaxy collisions and galaxy formation will dominate the dynamics of such systems, invalidating linear perturbation theory. Accordingly, although these modifications based on gravitational instabilities may prove useful in models of isolated disk galaxies, we choose not to include such modifications in our approach. In terms of the quiescent disk galaxy evolution, we note that in large part the effect of gravitational instability is to truncate star formation at large radius (typically several disk scale lengths). Using our model, typically more than 95% of the total star formation in isolated disk galaxy models occurs inside three disk scale lengths, implying that the lack of such a radial cutoff in our star formation law is of negligible importance to the overall properties of the galaxy. Of greater concern is the fact that our models ignore the possibility of suppression of star formation in the central regions of disks due to large galactic shear (e.g., Kenney, Carlstrom, & Young 1993). In our isolated disk models the central star formation rates contribute only modestly (i.e. $\sim 5\%$) to the global star formation rates, while in merger models which develop central starbursts (Mihos & Hernquist 1994a, 1994b), the central gas densities are much higher than inferred critical densities. Therefore for our purposes we feel a simple Schmidt law offers a reasonable parametrization of the star formation rate in galaxies.

Calculating feedback on the ISM is the next step in the calculation. Through both stellar winds from high mass stars and supernovae, star formation can inject a great deal of energy and newly formed metals into the surrounding ISM (see Tenorio-Tagle & Bodenheimer 1988 for a review). The deposited energy can be in both thermal and kinetic form; the relative fraction in each type will generally be a complicated function of the local ISM microphysics, but will be largely time-dependent, as the ISM rapidly thermalizes and radiates away the deposited energy. On small length and time scales, the deposited energy will be largely kinetic, in the form of shock fronts from supernovae and stellar winds; as these shock fronts expand, their kinetic energy will be transformed into thermal energy, such that on larger scales the energy deposition should be characterized by an increase in the thermal energy of the ISM. Therefore the mechanism by which this energy is transferred from star forming SPH particles to their nearby neighbors will be dependent on the length and timescales in question.

In previous implementations of star formation in SPH algorithms different approaches have been used to transfer this energy to the ISM. Katz (1992) used a Miller-Scalo (1979) initial mass function (IMF) to calculate the number of stars forming with mass $M > 8 M_\odot$. These stars were assumed to instantly form supernovae and release $10^{51}$ ergs of thermal energy each to the surrounding ISM. Summers (1993) also smoothed supernovae energy in purely thermal form, but using an exponentially decaying rate of transfer with an $e$-folding time of 10 Myr. Navarro & White (1993) took a different approach, smoothing a fraction of the supernovae energy as thermal energy, and the rest as a radial impulse given to the velocities of nearby SPH particles, thereby mimicking the kinetic energy input to the ISM. These studies revealed a common effect: since the thermal energy is input in regions of high density and short cooling time, the energy was rapidly radiated away and had little effect on the evolution of the system. The kinetic energy input by Navarro & White, however, could dominate the local dynamics of the gas depending on the relative (and adjustable) fraction of thermal and kinetic energy.

With these studies in mind, we adopt the following compromise for the injection of energy into the ISM. Following Katz (1992), we use a Miller-Scalo IMF to calculate the number of stars with mass $M > 8M_\odot$ and assume they instantly go supernovae and return their gas to the ISM. We then convert a fraction $\varepsilon_{kin}$ of the supernovae energy to kinetic energy by applying a radial kick to the velocities of nearby SPH particles in the following manner. First, SPH particles within two smoothing lengths of the star forming particle are identified. The smoothing kernel (see Hernquist & Katz 1989 for a detailed description of the smoothing kernel employed) is then used to assign each neighboring particle a fraction of the total kinetic energy injected by the star forming particle. In this manner, close neighbors receive a greater fraction of the injected energy than do distant neighbors, as might be expected for energy injection into the ISM. Each neighboring particle then receives a velocity impulse directed radially away from the star forming particle, with a magnitude $\Delta v_i = \sqrt{2(w_i \varepsilon_{kin} E_{SN})/M_i}$, where $i$ indicates the neighbor being perturbed and $w_i$ is the weighting based on the smoothing kernel. Using $\varepsilon_{kin} = 10^{-4}$ and $E_{SN} = 10^{51}$ ergs, $\Delta v_i$ is typically $\lesssim 0.1$ km s$^{-1}$ for nearby neighbors. The *total* amount of kinetic energy received by an individual particle (i.e. the sum of the contributions from all its neighbors) typically amounts to a perturbation of $\lesssim 1$ km s$^{-1}$, compared to the isothermal sound speed $c_s \sim 10$ km s$^{-1}$ at $10^4$ K. At each timestep, this process is repeated for all SPH particles, as each star-forming particle injects kinetic energy into the ISM.

In order to fix the amount of kinetic energy injected into the ISM via star formation, we constrain $\varepsilon_{kin}$ such that the ISM in an isolated disk model maintains a constant vertical scale height with time. If $\varepsilon_{kin}$ is too large, the isolated disk fattens with time as excessive energy is deposited into random motions of the ISM gas. Furthermore, star formation in the model disk is spatially very smooth and even, unlike the clumpy distribution of HII regions in spiral galaxies. Conversely, if $\varepsilon_{kin}$ is too small, then star formation provides too little support against fragmentation of the disk gas, resulting in an ISM which is overly prone to strong localized bursts of star formation. By adjusting $\varepsilon_{kin}$ in this way, we create a steady-state ISM which balances gravitational fragmentation with energy input from star formation. We ran a series of isolated disk models varying $\varepsilon_{kin}$ from $10^{-2}$ to $10^{-4}$ and found that values of $\varepsilon_{kin} > 10^{-3}$ resulted in a rapid increase in the vertical thickness of the disk gas. If $\varepsilon_{kin} = 10^{-3}$ thickening of the gas was less apparent, but small scale structures (i.e. HII regions) were still washed out. A value of $\varepsilon_{kin} = 10^{-4}$ gave good results in terms of the scale height of the disk gas and morphology of star formation (see §3.1). The remaining energy is considered to have been thermalized and radiated on short timescales (i.e. less than a single timestep, typically $\sim 10^6$ years). As noted in the previous work by Katz (1992), Summers (1993), and Navarro & White (1993), the results of such modeling are rather unaffected by thermal energy input from star formation, as such energy is radiated away almost instantaneously. Therefore we feel that our mechanism for energy injection provides for a reasonable qualitative description of the evolution of isolated disk galaxies.

The metallicity and gas mass return is handled in a straightforward manner. We assume instantaneous return of both gas and metals into the ISM from stars of mass $M > 8M\odot$. The mass return is calculated from the adopted IMF, assuming supernovae leave behind compact remnants of $1.4 M\odot$. For the Miller-Scalo IMF from 0.1–100 M$\odot$, 9% of the mass converted into stars at each time step is returned to the ISM. Metallicity return is characterized by the yield $y = M_{z,ret}/M_*$ (Tinsley 1980), where $M_{z,ret}$ is the total mass of all reprocessed metals and $M_*$ is the total mass locked up in stars. We use a yield $y = 0.02$ estimated both from observations of stellar metallicities in the Galaxy (Pagel 1987) and from theoretical calculations of mass return from high mass stars (e.g., Maeder 1992). We note that the actual details of this procedure will depend on the element in question, the time since the star formation event, and the form of the IMF; hence our prescription for metallicity return is only a simple approximation. However, given present uncertainties in the star formation law, IMF, and high-mass stellar yields, a more detailed version of this scheme is probably not yet well-motivated.

Once calculated, the local star formation rate must be applied to the SPH particles, depleting the ISM gas and forming a population of young stars. One common approach to this task is to create new stel-lar particles, which are given positions and velocities drawn from the parent SPH particle, and to then evolve these new star particles as collisionless particles (e.g., Katz 1992; Elmegreen & Thomasson 1993; Steinmetz & Müller 1994). However, for our purposes this procedure is unsatisfactory due to the physical and temporal resolution we wish to achieve. At any given timestep only a small fraction of the disk gas is converted into stars; furthermore, this star formation is spread throughout all the SPH particles. Therefore, for typical runs involving tens of thousands of SPH particles evolved for many dynamical times, more than $10^7$ new particles would need to be created and evolved, requiring a prohibitive amount of CPU time. Furthermore, the mass of such newly-created particles would be several orders of magnitude smaller than that of the surrounding particles representing the disk and halo of the galaxy; the subsequent scattering of the young star particles by the more massive disk and halo particles would result in a rapid, unphysical heating of the young population. In order to avoid the drawbacks associated with creating new particles, Summers (1993) employed a scheme in which particles were converted *in toto* from SPH to stellar particles. The disadvantage to this method lies in the fact that it sets the minimum resolution for a star forming event as the mass of an individual particle, which is typically $10^6$ M$_\odot$ for our models. Again, for detailed resolution of the star forming properties of individual galaxies, such methods are clearly undesirable.

In response to these computational issues, we take a new approach in modeling the young stellar population and its parent gas. We treat the SPH particles as hybrid gas/young star particles, characterized by both a *total* mass and a *gas* mass. The gravitational forces on these particles are calculated based on their total mass, while the hydrodynamical forces and properties are calculated using their gas mass. As an SPH particle forms stars its gas mass is reduced while its total mass stays fixed, thereby describing gas depletion due to star formation while still following the subsequent evolution of the newly formed stars, whose mass is simply given by $M_{i,*} = M_{i,tot} - M_{i,gas}$. When the gas mass fraction of an SPH particle drops below 5%, it is converted to a pure collisionless particle, with its remaining gas mass smoothed out among its nearest neighbors. Thus, these converted particles, along with the stellar mass fraction of the hybrid SPH particles, act as a good tracer of the young stellar population.

This method presents a number of computational advantages over methods which create new particles or convert existing particles completely. The total number of particles remains fixed, eliminating the need to spend computer resources evolving a large number of young star particles. In fact, as particles convert from hybrid SPH/young star particles to fully collisionless particles, the efficiency of the code actually improves, as less time is spent on the hydrodynamical portion of the calculation. Unlike methods which require complete conversion of some or all of the gas in an SPH particle into stars at a given timestep, our algorithm places no minimum constraint on the conversion amount, allowing the models to describe the mass and time scales necessary to examine the detailed star forming properties of the system.

The main disadvantage to our approach is that it assumes the newly formed stars and the parent gas are kinematically coupled until the gas is fully depleted. In reality the two components should evolve separately, since the gas will feel hydrodynamic forces and can dissipate energy, while the young stars will evolve in a collisionless manner. The discrepancy imposed by our procedure will be strongest if a dissipational event such as a strong shock occurs when a hybrid SPH/young star particle is composed of roughly equal amounts of gas and stars. In such cases, both the collisionless and hydrodynamical evolution will be compromised to some extent. Additionally, our technique will not be effective at modeling systems where gas depletion is a slow process (such as the evolution of disk galaxies over many Gyr), as the errors introduced by coupling the evolution of gas and young stars will continually grow. While our models for star formation, energy deposition, and metallicity enrichment are potentially very useful for studying the long term evolution of disk galaxies, in practice modifications will be necessary in order to decouple the evolution of the forming stars and their parent gas. Such modifications will probably involve a return to techniques which create new particles and therefore require a significant increase in the computational expense.

However, our models should be much better suited to modeling systems in which the conversion of gas to stars is a rapid process (i.e. starburst galaxies). In such circumstances, the transition of the hybrid SPH/young star particles from gas-dominated evolution to stellar-dominated evolution will be very rapid, significantly reducing the mass-weighted error of the calculation. The young stars formed in the starburst event will rapidly decouple from the gasdynamical evolution of the system, allowing their subsequent evolution to be accurately modeled. For this reason we feel the errors introduced by our method will be relatively minor for many of the applications we plan to explore with these models. In galaxy interactions and mergers, the star formation rates are rather modest until after the gas has dissipated its energy and is confined to small dense regions in the centers of the galaxies (Mihos et al. 1992, 1993; Mihos & Hernquist 1994a, 1994b). Typical gas depletion times in these starbursting galaxies are short, typically less than a few $\times 10^8$ years (e.g., Scoville & Soifer 1991; Stanford et al. 1990; Sage, Loose, & Salzer 1993), implying a rapid conversion of gas to young stars in these systems.

## 3. Galaxy Models

In this section, we use our star forming algorithms to model the dynamics of two different physical systems. The first is an isolated disk galaxy evolving quiescently. We use the properties of this system to constrain model parameters such as the initial star formation rate and energy injection into the ISM. We then turn to an interacting system to examine the star forming properties of ring galaxies.

### 3.1. Isolated Disks

Our galaxy models consist of a self-gravitating disk and halo, constructed using the technique of Hernquist (1993). The disk component follows an exponential mass profile:

$$\rho_d(R, z) = \frac{M_d}{4\pi h^2 z_0} \exp(-R/h) \text{sech}^2(\frac{z}{z_0}), \quad (4)$$

where $M_d$ is the disk mass, $h$ is the radial scale length, and $z_0$ is a vertical scale thickness. In the units employed here, where the gravitational constant $G = 1$, the disk has exponential scale length $h = 1$, a stellar vertical scale height $z_0 = 0.2$, and total mass $M_d = 1$. The halo is represented by a truncated "isothermal sphere" having a density profile

$$\rho_h(r) = \frac{M_h}{2\pi^{3/2}} \frac{\alpha(\gamma, r_c)}{r_c} \frac{\exp(-r^2/r_c^2)}{r^2 + \gamma^2}, \quad (5)$$

where $M_h$ is the total halo mass, $\gamma$ is a core radius, $r_c$ is a cutoff radius, and $\alpha(\gamma, r_c)$ is a normalization constant. In the models described here, the halos have total mass $M_h = 5.8$, scale length $\gamma = 1$ and cutoff radius $r_c = 10$. The combined disk and halo mass distribution yield a rotation curve which rises smoothly between $0 < R < 2$ and stays reasonably flat from $2 < R < 6$ (see Figure 1). The gas, comprising 10% of the disk mass and given an isothermal temperature of $10^4$ K, is distributed according to an exponential profile with a scale length identical to that of the stellar disk, but with a smaller vertical scale height. The total number of collisionless particles used to represent the disk and halo is $N_d = 32768$ and $N_h = 32768$, while the gas is represented by 16384 hybrid SPH particles. Scaling the model units to values typical of the Milky Way (e.g., Bahcall & Soneira 1980; Caldwell & Ostriker 1981), unit mass is $5.6 \times 10^{10}$ M$\odot$, unit length is 3.5 kpc, and unit time is $1.3 \times 10^7$ years.

The models are evolved using a system timestep of $\Delta t = 0.16$, using a tolerance parameter $\theta = 0.7$ and quadrupole moments to calculate the gravitational forces. Particles are assigned different gravitational softening lengths: $\epsilon_{halo} = 0.37$, $\epsilon_{disk} = \epsilon_{gas} = 0.08$. To calculate the hydrodynamic properties of the gas, a variable smoothing length is used such that each SPH particle has $N_s = 96$ neighbors within two smoothing lengths. Typical smoothing lengths are $\sim 0.1$ in the central regions of the disk, dropping to $\sim 1$ at large radius, where the particle density is low. Each SPH particle is allowed to have its own timestep in order to satisfy the Courant condition, with Courant number $C = 0.5$. Finally, the star formation calculations are performed using $C_{sfr} = 0.02$, $\varepsilon_{kin} = 10^{-4}$, and yield $y = 0.02$.

The evolution of the global star formation rate (SFR) and gas mass in an isolated disk galaxy is shown in Figure 2. The star formation rate is smooth and relatively constant over the time period shown (corresponding to $\sim 1$ Gyr), and only a small decrease in gas mass results from the consumption of disk gas. The $\sim$20% increase in the SFR at early times ($0 < T < 10$) is due to fragmentation of small clumps of gas from the smooth initial conditions. This fragmentation is quickly halted by the increased star formation in these denser regions, and the system quickly settles down to a steady-state evolution. In effect, then, the feedback in our models acts to harden the equation of state on small size scales. Figure 3 shows the evolution of the star-forming morphology of the system. As is typical in $N$-body simulations, the disk develops transient spiral features due mainly to swing amplification of noise in the potential (Toomre 1981; Hernquist 1993); the higher gas densities along these features result in correspondingly larger star formation rates as well. In a global sense, the models are stable over this time period, with no radial gas inflows or spontaneous starbursts developing. At the late stages of the evolution there is evidence for a growing bar instability in the disk due again to swing-amplified noise in the potential (Sellwood 1989), making models with much larger $N$ necessary to explore the evolution of disks on timescales greater than a few Gyr.

The "observed" star formation law is shown in Figure 4. To construct this plot, regions of the galaxy 0.15 units (500 pc) in radius were randomly sampled, measuring the volume density of gas and star formation at each point. The result plot shows a good approximation of a Schmidt law of index $n \sim 1.5$. The degree to which this "observed" Schmidt law will match the intended $n = 1.5$ law will depend on the size of the volume examined. As long as this size is comparable to the smoothing length on which the SPH estimate of gas density is based, good agreement between the intended and observed laws will be achieved. Since typical smoothing lengths in the inner disk are $\sim 0.1$ length units, similar to the radius of the observing volume, the observed law is very well described by the intended Schmidt law. At low gas densities the dispersion in the relation is larger, due to the lower particle densities and larger smoothing lengths in the outer regions of the disk.

The overall gas depletion time of the disk, as measured by $\tau_d = M_{gas}/\dot{M}_{gas}$, is 425 time units, or 5.5 Gyr, similar to depletion times inferred for nearby disk galaxies (Kennicutt 1983). A detailed map of gas depletion timescale over the disk is shown in Figure 5, and shows gas depletion times ranging from 100 time units (1.3 Gyr) in the center to >1000 time units (13 Gyr) in the outer disk. While central depletion times of 1 Gyr may seem worrisome in modeling evolution over this timescale, for Schmidt laws of index $n > 1$ the depletion times get *longer* over time, as star formation is self-quenched. For this reason, the actual central depletion of gas occurs over a much longer period than that described by the instantaneous gas depletion timescale. To illustrate this effect, Figure 6 shows the time evolution of the gas surface density, which displays only a 30% decline in the central region over 1 Gyr. While infall of primordial gas over long timescales may act to replenish the ISM and lengthen the gas depletion times, this slow infall would play only a minor role in the evolution of galaxies on the short timescales covered by our models.

Finally, the radial ISM metallicity profiles of the star forming disk are shown in Figure 7. The first panel shows the actual mass of heavy elements returned to the ISM, while the second panel shows the evolving metallicity profile of the disk, assuming an initial metallicity gradient of $-0.05$ dex kpc$^{-1}$, normalized to the solar value at the solar radius (see, e.g., Belley & Roy 1992 and references therein). Because of the density-dependent Schmidt star formation law and exponential profile of disk gas, the distribution of injected metals follows an exponential radial profile. As a result, the overall metallicity profile steepens with time. The metallicity profile of the *stellar* distribution is more complicated to construct, depending on the detailed star formation rate, evolving ISM metallicity, and metallicity of the background disk population. By keeping track of the evolving star formation rate and injected metal mass for each SPH/young star particle, we can construct the stellar metallicity profile at any time for a range of initial gas and stellar metallicities. Such a technique should prove powerful in probing the formation of radial gradients in both spiral and elliptical galaxies.

### 3.2. Ring Galaxies

Having established that our models mimic the evolution of isolated disk galaxies reasonably well (ignoring infall), we now turn to a model of an interacting pair of galaxies which gives rise to features like those in the prototypical ring galaxy AM 0035–35, the Cartwheel. The model is taken in large part from Hernquist & Weil (1993; hereafter HW), with minor modifications here. The primary (target) galaxy consists of an "isothermal" dark halo, and exponential stellar disk, and a hybrid exponential/flat distribution of ISM gas (see below). The mass ratio of these components is 2.9:0.9:0.3 respectively. Although the gas mass fraction of the disk is very high (33%), most of the gas is located at large radius, such that inside 2.5 disk scale lengths the gas represents only 10% of the disk mass. This large total gas mass is chosen in order to more accurately model the evolution of the ring at large radii, where the gas mass fraction of the disk may be very high. The penetrating satellite galaxy is modeled as a spherical, gas-free galaxy with a density profile (Hernquist 1990):

$$\rho_s(r) = \frac{M_s}{2\pi} \frac{a_s}{r} \frac{1}{(r+a_s)^3} \quad (6),$$

where $M_s$ is the satellite mass and $a_s$ is the radial scale length. The companion has mass $M_s = 1.0$, roughly 1/4 the primary galaxy mass, and scale length $a_s = 0.5$. The primary galaxy consists of collisionless particles which represent the stellar disk ($N_{disk} = 24576, \epsilon_{disk} = 0.08$) and dark halo ($N_{halo} = 32768, \epsilon_{halo} = 0.37$), and 12288 gas particles ($\epsilon_{gas} = 0.08$) which represent the ISM. The companion galaxy consists of 4096 collisionless particles ($\epsilon_{companion} = 0.1$. All components are fully self-gravitating. The parameters of the calculation are similar to those in the previous section, with the exception that the number of neighbors used in the SPH calculations has been reduced to $N_s = 30$ in response to the lowered number density of SPH particles in the disk.

The radial distribution of gas in the primary galaxy is somewhat different from the pure exponential profile used by HW, motivated by the weakness of the outer ring in their simulation. Observations of disk galaxies show significant amounts of gas at large radius, which may lead to surface density enhancements in ring galaxies in excess of those shown in the simulation by HW. To improve the likelihood of producing a strong outer ring, we increase the amount of gas at large radius, where the outer ring dominates the light profile. Specifically we use an exponential distribution of gas inside $R = 2.5h$, joining smoothly in surface density to a constant surface density distribution of gas from $R = 2.5h$ to $R = 8h$. Inside $R = 2.5h$, the ratio of gas to total surface density, $\Sigma_{gas}/\Sigma_{disk}$, equals 0.1, whereas outside $R = 2.5h$, $\Sigma_{gas}/\Sigma_{disk}$ is a monatomically increasing function, reaching a value of 1.0 at $R = 7h$. While this choice for the initial gas distribution is rather *ad hoc* and is chosen to increase the strength of the induced outer ring, it is not unlike the observed distribution of gas in many galaxies, which show an inner exponential distribution of CO gas and an outer, less-rapidly declining distribution of HI (Young & Scoville 1991). We choose the truncation radius for the gas distribution such that it is larger than the maximum radius of the propagating ring. Tests of this model in isolation show little evolution over $\sim 500$ Myr, the duration of the ring galaxy encounter, other than some slight outward diffusion of gas at the outer edge of the extended disk.

Figure 8 shows the evolution of the ISM component of the primary galaxy during the head-on collision. The morphology is nearly identical to that displayed in the simulation of HW, with the exception of the outer ring, which is now much more prominent due to the higher gas content of the outer disk. After passage of the companion through the disk ($T \sim 14$), the disk develops an outwardly propagating ring, due to radial oscillations in the orbits of the gas particles (i.e. Lynds & Toomre 1976). Unlike the ring observed in the stellar distribution, however, the gaseous ring is a true physical structure built up through shock dissipation and gas self-gravity (HW). By $T = 20$ the inner parts of the gas ring have fragmented and begin to fall back towards the center of the disk. As they fall back, these fragments shear out and form radial structures, or "spokes." A second ring also begins to expand out from the center late in the simulation.

An edge-on view of the encounter is shown in Figure 9, where the three dimensional nature of the disk response can now be seen. The passage of the companion through the primary disk excites vertical oscillations in the disk material, significantly thickening the outer disk. By the time the ring has propagated to $R = 5h$, the vertical thickness of the gas is quite large, $<z^2> \sim 2$ kpc. As a result, the outer ring is more correctly an encircling "ribbon" of material, rather than a compressed ring. Accordingly, the strong surface density enhancement observed in the face-on views corresponds to a relatively weak enhancement of volume density in the outer ring material. This strong disparity between surface and volume density in the ring suggests that strictly two dimensional simulations of ring galaxies may greatly overestimate the compression of disk gas in the ring.

The star forming properties of the ring galaxy during the encounter are illustrated in Figure 10, which shows the evolution of the global star formation rate, and in Figure 11, which depicts the morphology. The period of maximum star formation occurs late in the encounter, after the companion has pulled away from the primary, and is due not to the expanding outer ring, but rather the fragmentation of gas in the inner ring. The relatively low volume densities of gas in the outer ring – again, due to its vertical thickness – preclude extreme star formation in this region. The burst of star formation in this encounter is only a factor of 2–3 stronger than the initial rate, and consumes only a small amount of the disk gas. This relatively modest enhancement of star formation activity echoes the findings of Mihos et al. (1992), who found that signif-icant starbursts were achieved only in strongly interacting or merging galaxies. In a comparative study of the IRAS properties of isolated disks and interacting ring galaxies, Appleton & Struck-Marcell (1987) found that, on average, ring galaxies show comparable dust temperatures and only moderately enhanced far-infrared fluxes when compared to isolated disks. The relatively mild star formation enhancement suggested by that study compares well to the model described here.

Ring galaxies may be useful in understanding star formation because they are thought to have a simple history and geometry. A burst of star formation propagating outwards from the central part of the galaxy can set up radial age gradients in the stellar population of the disk. Indeed, radial color mapping of the Cartwheel in the optical and near infrared (Marcum, Appleton, & Higdon 1992) shows a trend for the disk colors to become bluer at large radius, which may reflect an age gradient due to star formation in the expanding ring. However, in light of the complicated evolution of the star forming gas, the formation of such simple radial gradients may not be easily achieved. Figure 12 shows the evolving radial distribution of star formation during the collision. A wave of star formation can clearly be seen moving out in radius after the collision, and is associated with the expanding compression wave in the ISM. As this ring moves out, the star formation fades with time, and subsequent infall of gas behind the ring triggers strong star formation in the central regions of the disk. As a result, any well-defined radial age gradients will exist only for a short period of time, as the inner starburst quickly dominates the star formation in the outer ring.

To examine the star forming properties of the model in more detail, we use as examples two specific times in the evolutionary sequence. First we consider the model at an early phase, $T = 18$, or approximately 60 Myr after the collision, when the first ring is expanding through the disk. Second, we look at a later time $T = 30$, or 220 Myr after the collision, when the primary galaxy displays both an inner and outer ring along with a system of radial spokes. It is at this time that the model most closely resembles the Cartwheel.

At early times, the star formation in the model is dominated by the outwardly propagating ring (Figures 11 and 12). As this ring moves out in radius, gas is swept up, leaving a region behind the ring devoid of both gas and star formation activity. The velocity structure in the disk gas (Figure 13a) shows a strong radial compression at the outer edge of the ring (R=2), with a very large velocity gradient in this region. Since this radial compression occurs in a region where the ISM is still in a thin disk ($<z^2> \sim 200$ pc), the compression is very dissipative, leading to a strong enhancement of gas density and star formation activity. Outside of this ring, the star formation properties are relatively unchanged from their preinteraction state, although there is some decrease in the level of star formation due to the increased vertical spread of the outer disk.

At later times, the conditions in the primary galaxy show more complicated structure. The velocity field (Figure 13b) shows three main regions: the outer ring, with an expansion velocity of 0.3 (80 km s$^{-1}$), the inter-ring region where material is falling back into the central region, and the inner ring, which shows an expansion velocity similar to that of the outer ring. At the position of the inner ring there is an extreme velocity gradient, which leads to strong shocking and dissipation of energy. It is along this inner ring that the gas fragments, leading to the strong star formation shown in the inner ring. The outer ring is now more diffuse, propagating into a region which is thick vertically ($<z^2> \sim 2$ kpc), resulting in a very mild compression of the gas and a relatively small increase in star formation activity.

In brief, the star formation evolution in this model may be characterized by two phases. First the galaxy experiences an early phase in which the first ring sweeps up the inner ISM, compressing the gas and forming stars along this expanding shock front. After this wave has traveled several radial scale lengths, the compressed gas falls back towards a second interior ring, and is subsequently reshocked in a second strong phase of star formation. Star formation in the outer expanding ring fades in intensity as the ring expands into regions of low surface density, which are thick vertically.

A comparison between the ring galaxy models and the Cartwheel yields mixed results. The morphology and expansion velocity of the ring at time T=30 match that of the Cartwheel (Fosbury & Harwarden 1979) quite well, suggesting that the Cartwheel is an older ring galaxy, with age $\sim 2 \times 10^8$ years. However, the star forming morphology of the Cartwheel, dominated by the outer ring with little trace of star formation in the inner regions, is more typical of our ring galaxy model at an earlier age of $\sim 5 \times 10^7$ years. We note that we have made no attempt to iterate on the model to reproduce the Cartwheel, and thus the Cartwheel's progenitor galaxies and orbit may be somewhat different from our model values. For example, the dearth of star formation in the central regions may be a result of gas being swept out by the passage of the companion, or simply indicative of a preexisting lack of central HI (see, e.g., Struck-Marcell & Higdon 1993). Furthermore, uncertainty in the masses of the Cartwheel's companion galaxies makes detailed modeling difficult. Although estimates of the masses of the companions (Davies & Morton 1982) suggest the mass ratio in the system is 10:1 (as opposed to the value of 4:1 used in our simulations), these estimates depend on the poorly-constrained dark matter content of the galaxies. We note, however, that the use of a less massive companion in our models would result in a much weaker response in the disk, resulting in a much *poorer* match to the rings and spokes in the Cartwheel (see HW for details). A detailed comparison of the observed star-forming morphology and two dimensional HI velocity field to those predicted by these models and the models of Struck-Marcell & Higdon (1993) would be useful in constraining the parameters of the encounter.

## 4. Summary

Modeling the complete evolution of individual galaxies has long proved difficult due to the coupled effects of gravitational stellar dynamics, hydrodynamics, and star formation. In particular, previous numerical models of disk galaxy evolution have been limited by the lack of a treatment of star formation and its effects on the different components of the galaxy. To address this deficiency, we have developed models of star forming disk galaxies which follow the full dynamical evolution of the stellar and gaseous components of the galaxy, with a detailed treatment of the effects of star formation. Starting with a hybrid $N$-body treecode and smoothed particle hydrodynamics code (TREESPH), we have included a treatment of star formation which models the depletion of gas, the formation of a young population of stars, and the injection of energy and metallicity into the surrounding ISM. We employ a Lagrangian form of a Schmidt law (Schmidt 1959) to calculate star formation rates in the model galaxies with spatial resolution of a few hundred parsecs. Gas depletion and the formation of new stars is handled through the use of hybrid gas/young star particles which slowly convert from SPH particles representing the ISM to pure collisionless particles which make up the newly formed stars. The energy input to the ISM from massive stars is treated as a kinematic "kick" to local gas particles, while the metallicity of the star forming gas is evolved using estimates of the heavy element yield from supernovae and stellar winds. By including a description of star formation and the evolution of the newly formed population, our models can, in principle, address issues such as the triggering of starbursts in galaxy interactions and mergers, and how these starbursts affect the kinematic and photometric properties of the evolving system.

Tests of the modeling technique using isolated disk galaxy models show the star forming properties to be stable over 1.5 Gyr. The modified Schmidt law used to calculate star formation yields a good match to a Schmidt law of index $n \sim 1.5$ The global star formation rate in the disk stays relatively constant, showing only a slow decline due to gas depletion. No instabilities or spontaneous starbursts are observed. The disk gas mass is reduced by 25% over 1.5 Gyr, and gas depletion times inferred from the model disk are comparable to those of nearby disk galaxies. Finally, the mass of injected metals is relatively small, and tends to slightly steepen any existing metallicity gradients in the disk gas.

Using the models, we simulate a galaxy collision which gives rise to a ring galaxy and find that ring galaxies go through two phases of star formation. Early in the evolution of the system, the galaxy is dominated by a ring of star formation induced by the compression of gas in an outwardly propagating density wave. Behind this ring of swept-up gas, little star formation occurs, while beyond the ring the star forming properties of the disk are relatively unchanged. As sufficient amounts of gas are swept up in the ring, the gas fragments and begins to fall back into the central regions, reshocking as it forms a second expanding ring. At this point the star formation rates are highest in the inner ring, while the star formation rates in the outer ring have declined due to the weak compression associated with material at large radii and large vertical scale height. This combination of declining star formation rates in the outer ring and the recompression of infalling gas acts to weaken the simple picture of radial age gradients in ring galaxies. As a result, detailed velocity maps of ring galaxies may prove more useful than broad band colors in probing the evolutionary history of these disturbed systems.

While our approach represents a significant advance in the modeling of disk galaxies, it is not without limitations. The use of hybrid SPH/young star particles assumes the gas and the newly formed stars are kinematically coupled until the gas is depleted, while in reality the two components will evolve separately. However, if these hybrid particles are predominantly in one phase or the other, the mass-weighted error will be small. Tests involving galaxy mergers show this condition to be valid (Mihos & Hernquist 1994a, 1994b). A further limitation of our approach is that it presumes the physical description of the ISM, as described by the equation of state, is similar in all regions of the galaxy. However, recent observations of our own Galaxy and nearby starburst galaxies suggest that the gas at the center of these systems is characterized by pressures which are 2–4 times the pressure in the local ISM (Spergel & Blitz 1992; Heckman, Armus, & Miley 1990). In essence, the equation of state describing the gas may be stiffer in the central regions of galaxies, perhaps due to the influence of magnetic fields (Spergel & Blitz 1992) or high cosmic ray fluxes (Suchkov, Allen, & Heckman 1993). Furthermore, given the differing ISM conditions in the central regions of galaxies, using a Schmidt law to characterize star formation in these regions may be inappropriate. Further simulations employing different equations of state and star formation prescriptions will be necessary to probe these hypotheses.

## 5. Acknowledgements

We thank the anonymous referee for a helpful referee's report which resulted in a significantly improved version of this paper. This work was supported in part by the Pittsburgh Supercomputing Center, the San Diego Supercomputing Center, the Alfred P. Sloan Foundation, NASA Theory Grant NAGW–2422, the NSF under Grants AST 90–18526 and ASC 93–18185 and the Presidential Faculty Fellows Program.

---

Fig. 1.— Rotation curve for the isolated disk/halo galaxy model, showing the contribution from the disk and halo.

Fig. 2.— Evolution of a) global star formation rate and b) disk gas mass in the isolated disk/halo galaxy model. Note the relative constancy of the star formation rate and the slow gas depletion rate; the gas depletion time for this model is $\sim 5.5$ Gyr.

Fig. 3.— Star formation maps of the isolated disk/halo galaxy model. Dark shading represents most intense star formation. The images are displayed with a logarithmic intensity stretch, and time is shown at the top of each frame.

Fig. 4.— Star formation as a function of density in the isolated disk/halo galaxy model. Also shown are lines corresponding to Schmidt laws of index $n = 1$ and $n = 2$.

Fig. 5.— Map of the gas depletion time in the isolated disk/halo galaxy model. Dark shading represents the shortest depletion times ($\sim 1$ Gyr in the disk center), while light shading represents long depletion time ($> 10$ Gyr in the outer disk).

Fig. 6.— Evolution of the disk gas mass distribution in the isolated disk/halo galaxy model.

Fig. 7.— a) Evolution of the mass of injected metals in the isolated disk/halo galaxy model. b) Evolution of the metallicity of the disk gas, assuming an initial metallicity gradient of $-0.05$ dex/kpc, normalized to [Fe/H]=0 at the solar radius.

Fig. 8.— Time evolution of the gas component in the ring galaxy model, seen face-on to the disk plane. Time is shown in the upper right; each frame measures 20 length units per edge.

Fig. 9.— Time evolution of the gas component and companion in the ring galaxy model, seen edge-on to the disk plane. Time is shown in the upper right; each frame measures 20 length units per edge.

Fig. 10.— Evolution of global star formation rate in the ring galaxy model in Figures 8 and 9.

Fig. 11.— Time evolution of the star forming morphology of the ring galaxy model in Figures 8 and 9. The images are displayed with a logarithmic intensity stretch, and time is shown at the top of each frame.

Fig. 12.— Time evolution of the radial distribution of star formation in the ring galaxy model in Figures 8 and 9.

Fig. 13.— Velocity distribution $(v_r, v_z)$ of the disk gas as a function of radius in the ring galaxy model in Figures 8 and 9. The velocity dispersion is shown by the vertical bars. a) T=18, b) T=30. Each point represents a fixed amount of gas mass.